\documentclass[12pt]{iopart}
\usepackage{graphicx}
\begin{document}

\title[Interference of Fano-Rashba conductance dips]{Interference of Fano-Rashba conductance dips}

\author{A. Renart$^{1,2}$, M. M. Gelabert$^{2}$ and L. Serra$^{1,2}$}

\address{$^1$ Institut de F´\'{\i}sica Interdisciplin\`aria i Sistemes Complexos IFISC (CSIC-UIB)
E-07122 Palma de Mallorca, Spain}
\address{$^2$ Departament de F\'{\i}sica, Universitat de les Illes Balears, E-07122 Palma de Mallorca, Spain}
\ead{llorens.serra@uib.es}
\begin{abstract}
We study the interference of two tunable Rashba regions 
in a quantum wire with one propagating mode. 
The transmission dips (Fano-Rashba dips) of the two regions
either cross or anti cross depending on the distance between 
the two regions. For large separations we find Fabry-P\'erot
oscillations due to the interference of forwards and backwards 
propagating modes. 
At small separations overlapping evanescent modes play a prominent role,
leading to an
enhanced transmission and destroying the conductance dip. Analytical expressions
in scattering-matrix theory are given and the relevance of the
interference effect in a device is discussed.
\end{abstract}

\submitto{\JPCM}
\maketitle

\section{Introduction}

The Rashba interaction is a spin-orbit coupling present in two-dimensional
electron gases (2DEG's) confined by asymmetric potentials in the perpendicular 
direction \cite{ras}. It has attracted a lot of attention, mostly due to its 
tunability by electrical gating \cite{nit,scha}. Indeed, a controlled spin-orbit
(SO) coupling offers exciting possibilities to manipulate electron spin and 
current in, so-called, spintronic devices \cite{zut}. A paradigm of 
spintronic device, the spin transistor suggested by Datta and Das \cite{dada},
relies on the Rashba-induced spin precession as an electron propagates in 
a semiconductor quantum wire. The feasibility of this working principle 
has been demonstrated in experiment only recently \cite{quay}.

Besides the constant-spin-orbit case, situations where the Rashba coupling
acting on a 2DEG is inhomogeneous in space have been theoretically addressed analyzing
interface-induced effects such as, e.g., spin accumulation, beam focussing 
and ``spin optics" \cite{kho,usa,mar,nik,glaz}. A finite SO region in a 2DEG has been shown 
to contain bound states purely induced by 
the spin-orbit coupling \cite{val}. In a quantum wire, a finite SO region 
produces quasibound states that quench the wire's conductance at specific
energies, i.e., dips appear in the conductance plateau for a given number
of propagating modes. In Ref.\ \cite{sase}, S\'anchez and Serra 
discussed how this mechanism
can be understood in terms of the well known Fano resonances of atomic 
physics \cite{fan}, suggesting
the name Fano-Rashba resonance for this conductance dips.
Fano-Rashba dips have been studied in presence of disorder \cite{shen,shen2} 
and under the influence of magnetic fields \cite{ssc}. 
Recently, a review on Fano resonances in nanoscale structures
has also been published \cite{mir}.

Our aim in this work is to study the interference of 
the Fano-Rashba conductance dips
of two sequential SO regions in a quantum wire, separated by a distance $d$
(see Fig.\ 1). Similar SO modulations, named Rashba superlattices, have been 
studied in Ref.\ \cite{Xu}. Independently tuning $\alpha_1$ and $\alpha_2$, 
the Rashba intensities of the two regions, 
the two conductance dips can be brought in closed proximity to each other.
We will show that 
for large separations $d$ the two dips can cross, while for small $d$'s an
avoided crossing of the dips is observed. This is reminiscent of the 
von Neumann-Wigner crossing rule of molecular levels \cite{bra}. In our case, 
the coupling is mediated by evanescent modes around each SO region. If $d$
is larger than the range of the evanescent modes, the dip-dip coupling vanishes
and a crossing behaviour is seen. On the other hand, for small $d$'s 
avoided crossing of the two dips is obtained when transport is enhanced
due to transmission from the first to the second region through evanescent modes.

The relevance of evanescent modes in confined (quasi-1D) transmission is well
known \cite{bag,cah,barb,evan}. For Dirac-delta impurities, Bagwell \cite{bag} showed
that the dependence of the transmission on the separation between scatterers
has two clear regimes: a) a Fabry-P\'erot regime for large separations 
where the dominant mechanism is the interference between 
forwards and backwards propagating modes between scatterers; b) at small
separations a regime where transmission occurs predominantly through
evanescent modes. This is precisely the physical scenario we have sketched 
above for the interference of two Fano-Rashba dips. It is also worth
stressing that transmission through
evanescent modes between scatterers has been 
proved relevant for the 
Anderson localization of disordered wires \cite{cah}.

In this work we will present numerical calculations of a quantum wire's
transmission in the presence of two tunable Rashba regions. The physical 
analysis of the relevant mechanisms will be performed using scattering 
matrix theory, composing the matrices of successive scatterers. As in Ref.\
\cite{cah}, we have considered a generalized formulation of 
scattering-matrix theory where propagating and evanescent modes are 
treated on an equal footing. Truncating to different numbers of evanescent 
modes we study quantitatively their relevance, as a function of the 
distance between the two Rashba regions. For the case of only one 
propagating mode, we obtain an analytical formula for the 
transmission. Finally, the use of sequential Rashba regions 
as a spin-orbit-controlled device is discussed.

\begin{figure}[t]
\centerline{\includegraphics[width=5.5cm,clip]{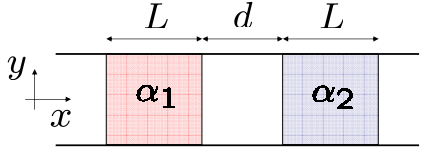}}
\caption{(Color online)
Sketch of the quantum wire with two tunable Rashba regions of length
$L$ separated by a distance $d$.}
\label{fig1}
\end{figure}

\section{Physical system and model}

We consider a 2DEG with a parabolic confinement along $y$ and free motion 
along $x$ described by the Hamiltonian
\begin{equation}
\label{eqh0}
{\cal H}_0 = \frac{p_x^2+p_y^2}{2m}+ \frac{1}{2}m\omega_0^2y^2\; .
\end{equation}
An inhomogeneous  Rashba interaction of type
\begin{equation}
{\cal H}_R = \alpha(x)\left(p_x\sigma_y-p_y\sigma_x\right)-i\frac{\hbar}{2} \alpha'(x)\sigma_y\, ,
\end{equation}
is active in the quantum wire. The Rashba intensity $\alpha(x)$ is assumed to vanish everywhere except in 
two separate regions where it takes the constant values $\alpha_1$ and $\alpha_2$. 
A sketch of the physical system is given in Fig.\ 1.
More precisely,
\begin{equation}
\label{eqalfa}
\alpha(x)=\alpha_1 {\cal F}_{x_1,L}(x) + \alpha_2 {\cal F}_{x_2,L}(x)\; ,
\end{equation}
where 
\begin{equation}
\label{eqf}
{\cal F}_{x_0,L}(x) = \frac{1}{1+e^{(x-x_0-L/2)/\sigma}} -\frac{1}{1+e^{(x-x_0+L/2)/\sigma}} 
\end{equation}
describes a square barrier of length $L$ centered at $x_0$. In Eq.\ (\ref{eqf}) the length $\sigma$ is 
introduced to model smooth space transitions with $\sigma\ll L$. The distance between the two
Rashba regions defined by Eq.\ (\ref{eqalfa}) is $d=x_2-x_1-L$ and it is always assumed
$d>0$ to avoid overlapping. Experimentally, the Rashba interaction can be controlled with 
gate electrodes modifying the $z$-asymmetry of the quantum well hosting the 2DEG \cite{nit,scha}. 
Our model would thus require
an independent tuning of the gates defining $\alpha_1$ and $\alpha_2$. Notice also that no electrostatic 
in-plane effects, other than the lateral potential $m\omega_0^2 y^2/2$ are contained in the model.

The transverse modes of Eq.\ (\ref{eqh0}) are characterized by
\begin{equation}
\begin{array}{rcl}
\left(\frac{p_y^2}{2m}+\frac{1}{2}m\omega_0^2y^2\right)
\phi_n(y) &=& \varepsilon_n \phi_n(y)\;,\\
\varepsilon_n &=& \left(n-\frac{1}{2}\right)\hbar\omega_0,\quad (n=1,2,\dots)\;.
\end{array}
\end{equation}
The 2D electron wave function $\Psi(x,y,\eta)$ where $\eta=\uparrow,\downarrow$ is the spin variable, 
fulfills the Schr\"odinger equation for a given energy $E$
\begin{equation}
({\cal H}-E)\Psi(x,y,\eta) = 0\; .
\end{equation}
As in Ref.\ \cite{sase} we expand the wave function in transverse and spin eigenmodes
\begin{equation}
\Psi(x,y,\eta)=\sum_{n=1,2,\dots,s=\pm}{\psi_{ns}(x)\phi_n(y)\chi_s(\eta)}\; ,
\end{equation}
where $\chi_s(\eta)$ are eigenspinors in $y$ direction. Projecting we find the equation
for each channel amplitude $\psi_{ns}(x)$, 
\begin{eqnarray}
\label{eq8}
-\frac{\hbar^2}{2m}\psi''_{ns}(x)
+(\varepsilon_n-E)\psi_{ns}(x)
+\sum_{ns}{\langle ns|{\cal H}_R|n's'\rangle \psi_{n's'}(x)}
&=& 0\; .
\end{eqnarray}
The matrix element of the Rashba interaction $\langle ns|{\cal H}_R|n's'\rangle$ in the 
$y\eta$ space is the only source of interchannel coupling. More specifically, the 
$p_x\sigma_y$ contribution to ${\cal H}_R$ is fully diagonal and only the $p_y\sigma_x$ 
induces
a coupling between $\psi_{ns}(x)$ and the splin-flipped neighbouring bands
$\psi_{n\pm1\bar{s}}(x)$. Next section contains the numerical results by solving
the system of coupled equations (\ref{eq8}) with the quantum transmitting 
boundary method. The reader is 
addressed to Ref.\ \cite{qtbm} and references therein for more details on the
numerical algorithm.
We will consider one propagating mode, $\varepsilon_1<E<\varepsilon_2$, and focus our attention
on the system conductance, determined by the quantum transmission with the help of
Landauer formula $G=T e^2/h$, where $T$ is the total quantum transmission obtained after
summing the modulus squared of the transmission amplitudes for the two 
spin channels $T=\sum_{ss'}{|t_{1s,1s'}|^2}$.

\section{Results}

\subsection{Numerical}

This subsection presents the transmission of the system obtained numerically with the method
of Ref.\ \cite{qtbm}. The total number of modes, both propagating and evanescent, in the 
linear system of equations (\ref{eq8}) is taken 
to be large enough to yield converged results. 
We focus on the Fano-Rashba conductance dips for a fixed $\alpha_1$ and 
varying $\alpha_2$. Dark regions represent the position of the conductance dips. The
figure clearly shows that for large separation between the two Rashba regions there is 
a crossing of the two dips that evolves to an anti crossing for small values. Remarkably, 
for an intermediate distance ($d=4\ell_0$) the two dips are in a perfectly 
destructive interference, leading to a high conductance at the position where the 
crossing would normally occur. We also notice that for very short distances the dips become
highly asymmetric, with one of them clearly dominating the other. The scenario presented
in Fig.\ 2 can be interpreted in terms of a $d$-dependent dip-dip coupling: vanishing
for large distances (crossing behaviour) and increasing at small $d$'s (anti crossing). 
We present in what follows evidence proving that the quantum wire evanescent modes 
mediate this coupling using, for this purpose, a
scattering matrix formalism.

\begin{figure}[t]
\centerline{\includegraphics[width=10cm,clip]{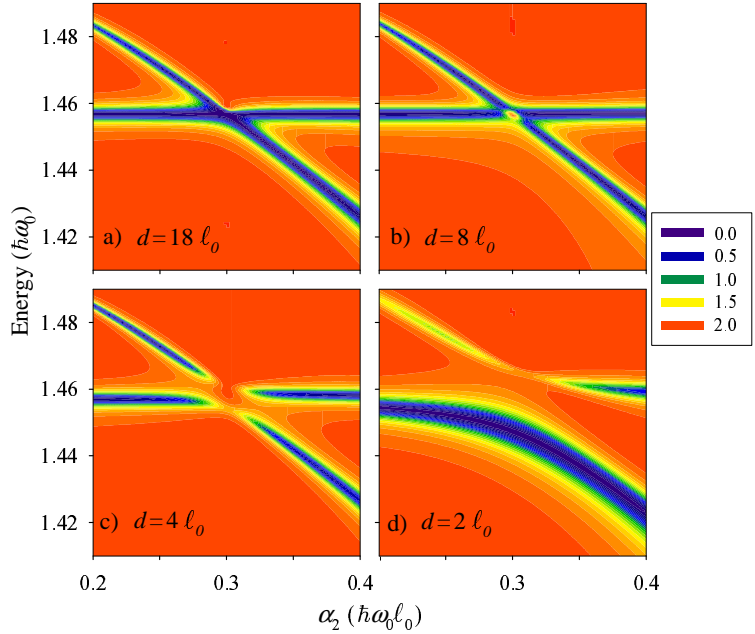}}
\caption{(Color online)
Conductance in units of $e^2/h$ as a function of $\alpha_2$ and $E$ for a fixed 
$\alpha_1=0.3\hbar\omega_0\ell_0$, $L=8\ell_0$ and $\sigma=0.1\ell_0$. Each panel corresponds to 
a different value of $d$, the distance between the two Rashba regions.
The wire parabolic 
confinement fixes our energy 
$\hbar\omega_0$ and length unit $\ell_0=\sqrt{\hbar/m\omega_0}$.
Dark (bright) colour indicates low (high) conductance.}
\label{fig2}
\end{figure}

\subsection{Scattering matrix theory}

Scattering phenomena in quantum mechanics with coherent 
wave functions are described by scattering matrix theory. For a single scatterer
there is a matrix of complex numbers relating the flux amplitudes of
outgoing channels $\{b_{c,ns}\}$ to those of 
incoming ones $\{a_{c,ns}\}$, where we introduced a ``contact" label $c=\ell,r$
(referring to left $\ell$ or right $r$), while $ns$ are indicating transverse mode and spin
as before. Namely, 
\begin{equation}
\label{eq9}
\left(
\begin{array}{c}
\sqrt{k_n}b_{\ell,ns}\\
\sqrt{k_n}b_{r,ns}
\end{array}
\right)
=
\left(
\begin{array}{cc}
r_{ns,n's'} & t'_{ns,n's'} \\
t_{ns,n's'} & r'_{ns,n's'} \\
\end{array}
\right)
\left(
\begin{array}{c}
\sqrt{k_{n'}}a_{\ell,n's'}\\
\sqrt{k_{n'}}a_{r,n's'}
\end{array}
\right) \; .
\end{equation}
As usual, a sum is implied for repeating indexes in Eq.\ (\ref{eq9}) and the 
factors $\sqrt{k_{n}}$ take into account the channel flux by introducing the 
channel wavenumbers
\begin{equation}
\label{eq10}
k_{n}=\frac{\sqrt{2m(E-\varepsilon_n)}}{\hbar}\; .
\end{equation}
The idea underlying scattering theory in quasi-1D transmission is that the 
wave function in the $c=\ell$ or 
$c=r$ regions, where the scatterer is no longer active, is given in terms 
of channel amplitudes and wavenumbers as
\begin{eqnarray}
\label{eq11}
\Psi_{c}(x,y,\eta) &=&
\sum_{ns}{
a_{c,ns}\,\phi_n(y)\chi_s(\eta)\, e^{is_c k_{n}(x-x_c)}
}\nonumber\\
&+&
\sum_{ns}{
b_{c,ns}\,\phi_n(y)\chi_s(\eta)\, e^{-is_c k_{n}(x-x_c)}
}\; .
\end{eqnarray}
In Eq.\ (\ref{eq11}) we have introduced the notation $s_\ell=1$ and $s_r=-1$
and $x_c$ is indicating the position  
where the scatterer becomes inactive for contact $c$.

For our present purposes, it is essential to realize that the number of channels
$\{ns\}$ in Eqs.\ (\ref{eq9}) and (\ref{eq11}) is, in principle, infinite \cite{cah}. 
For a 
given energy $E$ part of these channels will be propagating ($E\ge \varepsilon_n$)
and the rest will have an evanescent character. The intrinsic distinction between 
propagating and evanescent characters is that the wavenumber, Eq.\ (\ref{eq10}), 
is real in the former and purely imaginary in the latter. The physical meaning becomes
obvious when looking at the $x$-dependence of Eq.\ (\ref{eq11}). Though infinite, 
the number of evanescent channels is truncated in practice and fast convergence is usually 
obtained.

\subsection{Sequential scatterers}

Assuming the scattering matrix of one scatterer is known, the solution for two
identical scatterers can be obtained by adequately composing the matrices
of each scatterer. This 
procedure only requires to realize that the right output from the first scatterer
becomes left input for the second and vice versa. We need to label now the amplitudes 
with the ``impurity" index $i=1,2$ as $\{a^{(i)}_{c,ns}, b^{(i)}_{c,ns}\}$.
Assuming that all the input coefficients vanish except that of mode $n=1$
with spin $s_i$, 
$a^{(1)}_{\ell,1s_i}=1$, the linear system for the output coefficients reads
\begin{equation}
\label{eq12}
\left\{
\begin{array}{rcl}
b^{(1)}_{\ell,ns}
-\displaystyle\sum_{n's'}{t'_{ns,n's'}\, e^{ik_{n'}d}\, b^{(2)}_{\ell,n's'}}
&=& r_{ns,1s_i}\;,\\
b^{(1)}_{r,ns}
-\displaystyle\sum_{n's'}{r'_{ns,n's'}\, e^{ik_{n'}d}\, b^{(2)}_{\ell,n's'}}
&=& t_{ns,1s_i}\; ,\\
b^{(2)}_{\ell,ns}
-\displaystyle\sum_{n's'}{r_{ns,n's'}\, e^{ik_{n'}d}\, b^{(1)}_{r,n's'}}
&=& 0\; ,\\
b^{(2)}_{r,ns}
-\displaystyle\sum_{n's'}{t_{ns,n's'}\, e^{ik_{n'}d}\, b^{(1)}_{r,n's'}}
&=& 0\; .
\end{array}
\right.
\end{equation}

Equation (\ref{eq12}) can be viewed as a sparse linear system for the 
unknowns $\{b^{(i)}_{c,ns}\}$. It can be solved with standard sparse numerical
routines for a fairly large number of evanescent modes \cite{scho,har}.
Reversely, for just one propagating mode, or one propagating and one evanescent
mode, analytical solutions can be given that recover known results
for the composition of scatterers (see Appendix). Of all the output
amplitudes of Eq.\ (\ref{eq12}), we are  
interested in the total transmission amplitude $t_{1s_o,1s_i}\equiv b^{(2)}_{r,1s_o}$,
representing the right output from impurity 2 in channel $1s_o$ corresponding
to a left input in impurity 1 in channel $1s_i$, $a^{(1)}_{\ell,1s_i}=1$.

\begin{figure}[t]
\centerline{\includegraphics[width=8cm,clip]{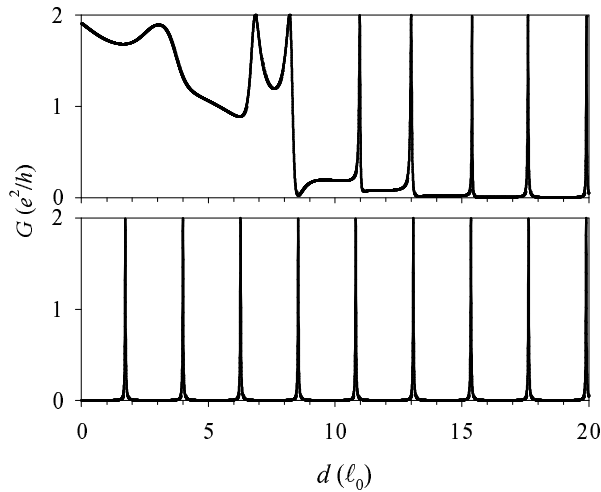}}
\caption{
Conductance as a function of distance $d$ between Rashba regions
for an energy $E=1.457\hbar\omega_0$ and $\alpha_1=0.3\hbar\omega_0\ell_0$
obtained with the method of scatterer composition.
Upper panel is the result including evanescent modes while lower panel
only considers the propagating mode.}
\label{fig3}
\end{figure}

The method of scatterer composition allows us to investigate 
the dependence on $d$, the distance between impurities, in 
an explicit way from Eq.\ (\ref{eq12}). A technical point worth 
of stressing is that an important simplification occurs for identical 
scatterers placed sequentially along $x$; namely, 
the scattering matrix is the same for each scatterer.
Figure 3 shows the result obtained as a function of $d$ for the energy 
and Rashba intensity of the conductance dip of Fig.\ 2. When evanescent modes
are fully neglected (lower panel) the transmission of the system
vanishes except for a sequence of very narrow, equally spaced peaks. 
They correspond to a Fabry-P\'erot-like regime \cite{bag} with 
constructive interference 
at distances such that an exact 
multiple of the electron wavelength fits in between Rashba regions.
This behavior changes dramatically for low distances 
when evanescent modes are included (upper
panel): the dip is effectively destroyed by evanescent-mode transmission
from the first to the second Rashba region. 
This effect exactly corresponds to the anti crossing seen in Fig.\ 2 
at small distances.
With the resolution of Fig.\ 3 upper panel, it is 
enough to include one evanescent mode, the contribution from higher ones 
being exceedingly small.

\subsection{Device}

The conductance dips discussed above are quite narrow and, therefore, not
robust against thermal or disorder fluctuations. Their observation 
requires the use of very low temperatures and purely ballistic samples.
It was shown in Ref.\ \cite{sase} that for stronger $\alpha$'s broader
dips are induced at the end of the first conductance plateau.
For a more robust conductance dip, in this section we analyze
the effect discussed in this paper in a device in which 
current is controlled by manipulating the intensity of successive 
Rashba regions (See Fig.\ 4). The idea that a superlattice of this type
could be of importance in practical application was already pointed
out in Refs.\ \cite{shen,shen2}. Our purpose here is to analyze this 
mechanism from the point of view of interference between Fano-Rashba dips
through evanescent modes.

Figure 4 displays the conductance for up to 3 regions with a strong
ratio $\alpha/\hbar\omega_0\ell_0$. For a single region there is
a sizeable dip which, however, does dot extend all the way to zero  (solid line).
Adding more regions at distance $d=2\ell_0$ has the effect 
to enhance the dip forming a quasi gap amenable 
to practical applications (lower panels).
It is remarkable
how for just two or three regions with $d=2\ell_0$ a quasi energy gap 
clearly develops at the dip position
$E\approx 1.25\hbar\omega_0$. At short distances 
the coupling through evanescent modes 
destroys the dip (upper panels) --notice, however, that a second narrow dip
appears at $E\approx 1.4\hbar\omega_0$  for two regions 
(dashed line, upper left panel)
but it is removed for 3 sequential regions (dash-dotted line).
A device based on the tuning of $\alpha$ for sequential Rashba regions
at a proper distance
would not  require the use of polarized leads, as compared to 
the Datta-Das spin transistor. Its basic shortcoming, however, is 
the sensitivity to the incoming electron energy which should lie in the 
region of the quasi gap. Increasing the number of sequential regions
makes the quasi gap more robust. The distance between Rashba regions
should be chosen appropriately in order to avoid destructive interference
through evanescent modes.

\begin{figure}[t]
\centerline{\includegraphics[width=10cm,clip]{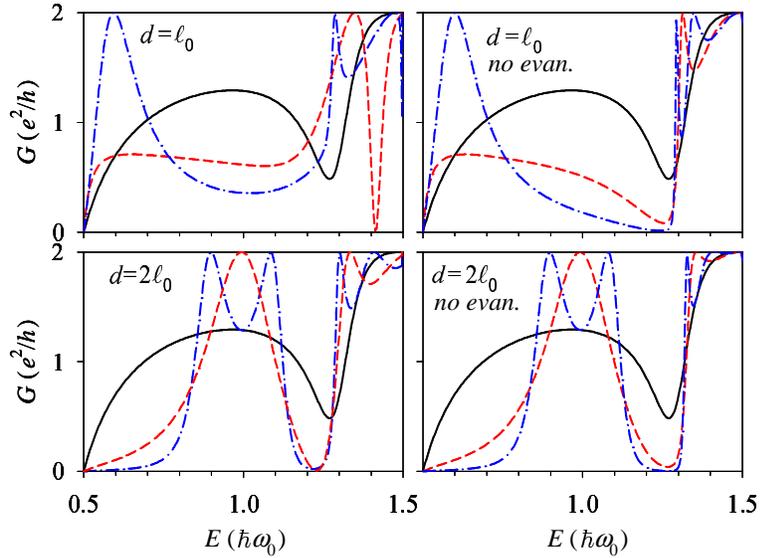}}
\caption{(Color online)
Upper: 
Conductance as a function of energy 
for sequential Rashba regions, each one having $L=\ell_0$ and 
$\alpha=\hbar\omega_0\ell_0$. The different curves correspond
to $1$ (solid), $2$ (dash) and $3$ (dash-dot) sequential regions. Upper and lower 
panels are for small and large separation $d$
between regions, respectively.  For comparison, right panels show the result
when evanescent channels are not included.
Lower panels suggest transistor operation 
by tuning $\alpha$, 
for $E/\hbar\omega_0\approx 1.25$,
with
the OFF and ON states represented 
by $\alpha/\hbar\omega_0\ell_0\approx 1$  and $\approx 0$, respectively.
}
\label{fig4}
\end{figure}

\section{Conclusions}

The interference of the Fano-Rashba dips of two successive Rashba regions in a quantum wire 
has been analyzed. As a function of the separation 
the two dips evolve from an anti crossing behaviour at large distances to a crossing
when the two regions are close. The physics has been interpreted in terms of a dip-dip 
coupling mediated by the wire's evanescent modes. The generalized formulation within scattering 
matrix theory, including evanescent and propagating modes on an equal footing,
has been discussed. The numerical solution of the resulting linear equation system has been 
implemented. In the limit of only one or two modes analytical expressions have been given. 
Finally, the application to a device in which current is controlled by tuning 
two or three sequential Rashba regions has been discussed. A main obstacle in practice is the energy sensitivity
of the Fano-Rashba dip. The conductance quasi-gap 
is destroyed at short distances and it becomes more and more robust when 
increasing the number of Rashba regions for a proper value of $d$. 

\ack
We thank David S\'anchez for many insightful discussions and suggestions.
This work was supported by grant No.\ FIS2008-00781 of MICINN (Spain).

\appendix

\section{Analytical}
For only two modes it is possible to obtain analytical solutions to the linear
system Eq.\ (\ref{eq12}). Let us assume there are only one propagating $n=1$ and one
evanescent $n=2$ modes. Taking into account spin, the set
of channels splits into two
coupled subsets $\{1+,2-\}$ and $\{1-,2+\}$. Since both subsets are equivalent
we restrict to the first one by considering incidence in mode $1+$. 
The transmitted output amplitude reads (spin indexes are not explicitly written
to simplify notation)
\begin{eqnarray}
\label{eqA1}
b^{(2)}_{r,1}
&=&
\frac{t_{11}t_{11}\, e^{ik_1d}}{1-R_{11}-\frac{R_{12}R_{21}}{1-R_{22}}}
+
\frac{t_{11}R_{12}t_{21}\, e^{ik_1d}}{(1-R_{11})(1-R_{22})-R_{12}R_{21}}
\nonumber\\
&+&
\frac{t_{12}t_{21}\, e^{ik_2d}}{1-R_{22}-\frac{R_{12}R_{21}}{1-R_{11}}}
+
\frac{t_{12}R_{21}t_{11}\, e^{ik_2d}}{(1-R_{11})(1-R_{22})-R_{12}R_{21}}\; ,
\end{eqnarray}
where we have defined
\begin{eqnarray}
\label{eqA2}
R_{n_1n_2}&=&r'_{n_11}r_{1n_2}e^{i(k_1+k_{n_2})d}
            +r'_{n_12}r_{2n_2}e^{i(k_2+k_{n_2})d}\; ,
\end{eqnarray}
with $n_{1,2}=1,2$. 

The explicit dependence on $d$, the distance between Rashba regions, is contained 
in Eqs.\ (\ref{eqA1}) and (\ref{eqA2}). To analyze the large-$d$ limit 
we recall that the evanescent wavenumber is purely imaginary 
$k_2=i\kappa_2$ ($\kappa_2>0$). As a result we get in that 
limit $e^{ik_2d}\to 0$
as well as $R_{12}\to 0$, $R_{22}\to 0$ and
\begin{eqnarray}
R_{11}&\to& r'_{11}r_{11}e^{i2k_1d}\nonumber\\
R_{21}&\to& r'_{21}r_{11}e^{i2k_1d}\; .
\end{eqnarray}
The transmitted amplitude is then 
\begin{equation}
\label{eqA4}
b^{(2)}_{r,1}=
\frac{t_{11}t_{11}\, e^{ik_1d}}{1-r'_{11}r_{11}e^{i2k_1d}}\; ,
\end{equation}
which is a familiar relation frequently used for single mode conductors. Equation (\ref{eqA1})
contains the analytical $d$-dependence that generalizes Eq.\ (\ref{eqA4}) in the 
presence of one evanescent channel. This causes, as shown in Fig.\ 3, 
a modification of the transmission resonances at short distances.

\end{document}